\documentclass[10pt,a4]{article}

\usepackage{latexsym,amsmath,amscd,amssymb,graphicx}

\begin{document}
%\begin{titlepage}
%
\title{Representations of the Poincar\'{e} group on relativistic phase space}%\\ (draft)}

\author{Yaakov Friedman\\
Jerusalem College of
Technology\\ P.O.B. 16031 Jerusalem 91160\\
Israel}
\date{}
\maketitle
\begin{abstract}

We introduce a complex relativistic phase space as the space
$\mathbb{C}^4$ equipped with the Minkowski metric and with a
geometric tri-product on it. The geometric tri-product is similar to
the triple product of the bounded symmetric domain of type IV in
Cartan's classification, called the spin domain. We define a spin 1
representations of the Lie algebra of the Poincar\'{e} group by
natural operators of this tri-product on the complex relativistic
phase space. This representation is connected with the
electromagnetic tensor. A spin 1/2 representation on the complex
relativistic phase space is constructed be use of the complex
Faraday electromagnetic tensor. We show that the Newman-Penrose
basis for the phase space determines the Dirac bi-spinors under this
representation. Quite remarkable that the tri-product representation
admits only spin 1 and spin 1/2 representations which correspond to
most particles of nature.



\end{abstract}
%
%\end{titlepage}
\section{Introduction}

The spin factor, a bounded symmetric domain of type IV in the Cartan
classification \cite{CE35}, can play an important role in physics.
It was shown in \cite{FR92} that the state space of any two-state
quantum system is the dual of a complex spin factor and the geometry
of the state space can be defined by the triple product of the spin
domain. In \cite{F04} and \cite{FS} it was shown that a new dynamic
variable, called
 $s$-velocity, which is a relativistic half of the usual velocity, is
useful for solving explicitly relativistic dynamic equations. It was
shown that the automorphism group, generated by $s$-velocity
addition, coincides with the conformal group. The Lie algebra of
this group is described by the triple product defined by the spin
domain.

In \cite{F07} we introduced a geometric tri-product as a
generalization of the geometric product in Clifford algebras. This
product coincides with the triple product of the spin domain. In
this paper, in order to make this mathematical model closer to the
physical reality, we modify this tri-product and define a
relativistic phase space as follows. We equip the space
$\mathbb{C}^4$  with an  inner product based on the Lorentz metric
and define a new geometric tri-product on it. This space is used to
represent both the space-time coordinates and the four-momentum of
an object. The real part of the inner product extends the notion of
the Lorentz scalar product on the space-time and energy-momentum,
while the imaginary part extends the symplectic skew scalar product
of the classical phase space.

We construct both spin 1 and spin 1/2 representations of  the
Poincar\'{e} group by natural operators of the tri-product on the
phase space. The generators of space-time translations of the Lie
algebra of the Poincar\'{e} group are represented by the basic
vectors of the relativistic phase space while the generators of
relativistic angular momentum are represented by natural operators
of the triple product. More precisely, for spin 1 representation,
the generators of boosts are presented by operators with the meaning
of electric field strength tensors while the generators of rotations
are presented by operators with the meaning of magnetic field
strength tensors. For the spin 1/2 representation,  the generators
of boosts and  of rotations are presented by the complex Faraday
electromagnetic tensor. We show that if we use the Newman-Penrose
basis on the phase space, we obtain the Dirac bi-spinors under this
representation.

In a forthcoming paper we will describe the significance of the
Newman-Penrose basis for the relativistic phase space.

\section{Commutation relations of the Poincar\'{e} algebra}

In this section we recall several known facts about the Lie algebra
of the Poincar\'{e} group. It is well-known that the laws of physics
must be invariant under the following transformations: 1) the
space-time translations, 2) the space rotations, and 3) the proper
Lorentz transformations (boosts). These transformations generate the
Poincar\'{e} group.

A basis of the Lie algebra of the Poincar\'{e} group consists of
generators of space-time translations, denoted by $P_\mu$ for
$\mu=0,1,2,3$ (all Greek indices range from 0 to 3) and generators
of rotations and boosts, called relativistic angular momentum,
denoted by $M_{\alpha\beta}$. For a scalar particle, described by a
wave function on flat space-time with the metric tensor $\eta
_{\mu\alpha}=diag(1,-1,-1,-1)$, these generators act as
\begin{equation}\label{gentranslation}
  P_\mu=\frac{\partial}{\partial x^\mu},
\end{equation}
and
\begin{equation}\label{angmomentdef}
 M_{\alpha\beta}=X_\alpha P_\beta-X_\beta P_\alpha ,
\end{equation}
where the operator $X_\alpha$ acts as multiplication of the wave
function by $x_\alpha=\eta _{\mu\alpha} x^\mu$.

The commutation relations of the Poincar\'{e} algebra in flat space
with the metric tensor $\eta _{\mu\alpha}$ are given by
\begin{equation}\label{comuttrans}
  [ P_\mu, P_\nu]=0,
\end{equation}
\begin{equation}\label{comutangtrans}
  [ M_{\alpha\beta},P_\mu]=\eta _{\mu\beta}P_\alpha-\eta
  _{\mu\alpha}P_\beta,
\end{equation}
and
\begin{equation}\label{Lortzcommut}
 [ M_{\mu\nu},M_{\alpha\beta}]=\eta _{\mu\beta}M_{\nu\alpha}+\eta _{\nu\alpha}
 M_{\mu\beta}-\eta _{\mu\alpha}M_{\nu\beta}-\eta _{\nu\beta}M_{\mu\alpha}.
\end{equation}
Recall also additional commutation relations:
\begin{equation}\label{dualitymatric}
  [P_\mu,X_\alpha]=\eta _{\mu\alpha}I,
\end{equation}where $I$ denotes the identity operator,
\[[X_\alpha,X_\beta]=0,\] and
\begin{equation}\label{comutangtransdual}
  [ M_{\alpha\beta},X_\mu]=\eta _{\mu\beta}X_\alpha-\eta _{\mu\alpha}X_\beta.
\end{equation}

The commutation relations (\ref{comuttrans})-(\ref{Lortzcommut}) of
the Poincar\'{e} algebra show that the real span of its generators
$\mathcal{L}=span_R \{P_\mu, M_{\alpha\beta}\}$ has a structure of a
graded Lie algebra. The $span_R \{M_{\alpha\beta}\}$ is a Lie
subalgebra $\mathcal{L}_0$ corresponding to the grade 0. This is the
Lie algebra of the Lorentz group as a subgroup of the Poincar\'{e}
group. The bracket of elements of $\mathcal{L}_0$ with $P_\mu$
belong to $span_R \{P_\mu\}$. So, $\mathcal{L}_1= span \{P_\mu\}$ is
the grade 1 of the algebra. Commutation relations (\ref{comuttrans})
imply that the brackets on $\mathcal{L}_1$ are trivial. So
\begin{equation}\label{LieAlgGradetReal}
   \mathcal{L}=\mathcal{L}_0\oplus\mathcal{L}_1,\;\;\mathcal{L}_0=span_R
\{M_{\alpha\beta}\},\;\mathcal{L}_1= span_R \{P_\mu\}
\end{equation}
is a graded Lie algebra with a representation of the Poincar\'{e}
algebra on it.

Our aim is to find a spin 1 and spin 1/2 representations of  the
Poincar\'{e} algebra. To do this we will complexify $\mathcal{L}$ by
introducing a complex relativistic phase space with a tri-product on
it.

\section{A complex relativistic phase space}

For description of classical motion of a point-like body, the
classical phase space, composed from space position and the
3-momentum, is used. A symplectic structure on the phase space
provides information on the scalar product and the antisymmetric
symplectic form of this space. For the classical phase space this
structure can be expressed efficiently by introducing the complex
structure and a scalar product under which the classical phase space
becomes equal to $\mathbb{C}^3$, see \cite{Kaiser} and
\cite{Arnold}.

 For description of the motion of a \textit{relativistic} point-like body we may use the
relativistic phase space. This is obtained by adding the time and
the energy variables to the classical phase space. So, the
relativistic phase space may be identified with $\mathbb{C}^4,$ in
which the real part expresses the 4-momentum and the imaginary part
represents the space-time position of a point-like body.

 To extend the symplectic structure to this
space, we will use the complex-valued scalar product introduced by
E. Cartan, see \cite{CE66}. Since the scalar product needs to
provide information on the interval of the 4-vectors, we replace the
Euclidian metric, used usually, with the Lorentzian one. We chose
the scalar product to be complex linear in both terms and not, as
usually, conjugate linear in one term. This is consistent with the
fact that the Lorentz invariant of an electro-magnetic field is
given by $\mathbf{F}^2=(\mathbf{E}+i\mathbf{B})^2$ and not by
$|\mathbf{F}|$.

 Since at any given point the action of generators of
space-time translations $P_\mu$ could be identified with four
vectors in the tangent space, we will present them by basic real
vectors $\{\mathbf{u}_\mu\}$ in $\mathbb{C}^4$. We define a scalar
product $<\cdot|\cdot>$ on $\mathbb{C}^4$ as follows:

\noindent {\bf Definition 3.1} \textit{ On the real basis vectors of
$\mathbb{C}^4$ a scalar product $<\cdot|\cdot>$ is given by
\begin{equation}\label{seslinearbasisdef}
  < \mathbf{u}_\mu|
  \mathbf{u}_\nu>=\eta _{\mu \nu}.
\end{equation}
 For two arbitrary vectors $\mathbf{a},\mathbf{b}\in\mathbb{C}^4$ it is given by
\begin{equation}\label{Minkbilin}
  <\mathbf{a}|\mathbf{b}>=<{a}^\mu \mathbf{u}_\mu|b^\nu
  \mathbf{u}_\nu>=\eta _{\mu \nu} {a}^\mu b^\nu,
\end{equation} where $\mathbf{a}=a^\mu
\mathbf{u}_\mu$ and $\mathbf{b}=b^\mu \mathbf{u}_\mu$}.

Evidently, this scalar product is bilinear and symmetric. For an
arbitrary element $\mathbf{a}\in \mathbb{C}^4$, the \textit{scalar
square} is given by
\begin{equation}\label{squaredef}
    \mathbf{a}^2=<\mathbf{a}|\mathbf{a}>=\eta _{\mu \nu} {a}^\mu
    a^\nu ,
\end{equation}
which is a complex number, not necessary positive or even real.

Note that the real subspace $M$ defined by the real vectors of
$\mathbb{C}^4$ with the bilinear form (\ref{squaredef}) may be
identified with the Minkowski space. The same is true for the pure
imaginary subspace $iM$ defined by pure imaginary vectors. We use
the subspace $M$ to represent the four-vector momentum
$\mathbf{p}=p^\mu \mathbf{u}_\mu$ and the subspace $iM$  to
represent the space-time coordinates $i\mathbf{x}=x^\mu i
\mathbf{u}_\mu$ of a point-like body in an inertial system. Thus,
the space $\mathbb{C}^4$ represents both space-time and the
four-momentum as \[\mathbf{a}=a^\mu\mathbf{u}_\mu \;\mbox{with}\;\;
a^\mu=p^\mu+ix^\mu.\]

 The square of elements of $M$ are
$\mathbf{p}^2$ which is known to be equal to $(m_0c)^2$ with $m_0$
the rest mass of the object, while the negative of the square
$-(i\mathbf{x})^2$ of elements of $iM$ have the meaning of
space-time interval. Note that both
$\mathbf{a}^2=\mathbf{p}^2-\mathbf{x}^2+2i<\mathbf{p}|\mathbf{x}>$
and $<\overline{\mathbf{a}}|\mathbf{a}>=\mathbf{p}^2+\mathbf{x}^2$
(where $\overline{\mathbf{a}}=\overline{a}^\mu \mathbf{u}_\mu$
denotes the complex conjugate of $\mathbf{a}$) are Lorentz
invariants and that
\[\mathbf{p}^2=\frac{1}{2}Re(<\overline{\mathbf{a}}|\mathbf{a}>+\mathbf{a}^2),\;
\mathbf{x}^2=\frac{1}{2}Re(<\overline{\mathbf{a}}|\mathbf{a}>-\mathbf{a}^2).\]

For any two vectors $\mathbf{a},\mathbf{b}\in \mathbb{C}^4$ we may
consider also a scalar product
$<\overline{\mathbf{a}}|\mathbf{b}>=\eta _{\mu \nu} \overline{a}^\mu
b^\nu.$ The Lorentzian scalar product is the real part
\begin{equation}\label{intervalascomplscalar prod}
    Re<\overline{\mathbf{a}}|\mathbf{b}>=\frac{1}{2}\eta _{\mu \nu}
(\overline{a}^\mu b^\nu+{a}^\mu \overline{b}^\nu),
\end{equation}
 which is symmetric and
 extends the Lorentzian product (and the notion of an interval) from the subspaces $M$ and $iM$ to
$\mathbb{C}^4=M\oplus iM$.

The imaginary part defines a skew scalar product
\begin{equation}\label{liebracL1}
    [\mathbf{a},\mathbf{b}]=Im<\overline{\mathbf{a}}|\mathbf{b}>=
    \frac{1}{2i}\eta _{\mu \nu}
(\overline{a}^\mu b^\nu-{a}^\mu \overline{b}^\nu),
\end{equation}
which extends the symplectic skew scalar product. This bracket can
be used to define the Poisson bracket of two functions and two
vector fields. Thus, the space $\mathbb{C}^4$ with the scalar
product (\ref{Minkbilin}) can be used as a basis for a relativistic
phase space.

The commutation relations (\ref{comutangtrans}) suggest that the
relativistic angular momentum $M_{\alpha\beta}$ can be presented as
operators on the space $\mathbb{C}^4.$ Moreover, such operators will
be bilinear in $\mathbf{u}_\alpha ,\mathbf{u}_\beta$ and
antisymmetric in these variables. The commutation relation could be
used to define a triple product of $\mathbf{u}_\alpha
,\mathbf{u}_\beta ,\mathbf{u}_\mu$.

\medskip

 \noindent \textbf{Definition 3.2} \textit{Let  $\mathbb{C}^4$ denote a
 4-dimensional complex space with the scalar
product (\ref{Minkbilin}). A }\textbf{geometric tri-product}\textit{
$\{\;,\;,\;\}:\mathbb{C}^4\times\mathbb{C}^4
\times\mathbb{C}^4\rightarrow \mathbb{C}^4$ is defined for any
triple of elements $\mathbf{a}$, $\mathbf{b}$ and $\mathbf{c}$  as
\begin{equation}\label{tripleproddef}
    \mathbf{d}=\{\mathbf{a},{\mathbf{b}},\mathbf{c}\}=<\mathbf{a}
    |{\mathbf{b}}>\mathbf{c}-<{\mathbf{c}}
     |\mathbf{a}> {\mathbf{b}}+<
    {\mathbf{b}}| \mathbf{c}> \mathbf{a}.
\end{equation}
}

In the basis $\{\mathbf{u}_\mu\}$ definition (\ref{tripleproddef})
takes the form
\begin{equation}\label{geomtriprodtensor def}
d^\mu=\eta _{\alpha\beta}a^\alpha {b}^\beta c^\mu-\eta
_{\alpha\beta}c^\alpha a^\beta {b}^\mu+ \eta
_{\alpha\beta}{b}^\alpha c^\beta a^\mu .
\end{equation}

\noindent \textbf{Definition 3.3} For any pair of elements
$\mathbf{a},\mathbf{b}\in \mathbb{C}^4$ we define a linear map
$D(\mathbf{a},\mathbf{b}):\mathbb{C}^4\rightarrow\mathbb{C}^4$ as
\begin{equation}\label{Ddef}
   D(\mathbf{a},\mathbf{b})\mathbf{c}=\{\mathbf{a},{\mathbf{b}},\mathbf{c}\}
\end{equation}
and an antisymmetric map $\hat{D}(\mathbf{a},\mathbf{b})$ as
\begin{equation}\label{Dhatdef}
   \hat{D}(\mathbf{a},\mathbf{b})=\frac{1}{2}(D(\mathbf{a},{\mathbf{b}})-
   D({\mathbf{b}},\mathbf{a})).
\end{equation}

It is easy to verify that the geometric tri-product satisfies the
following properties:

\noindent \textbf{Proposition 3.1} The tri-product, defined by
(\ref{tripleproddef}), satisfies:
\begin{enumerate}
 \item $ \{\mathbf{a},{\mathbf{b}},\mathbf{c}\}$ is complex linear in
 all variables
$\mathbf{a},$ $\mathbf{b}$ and $\mathbf{c}.$
\item The triple product is symmetric in the pair of outer variables
\begin{equation}\label{symtriple prod}
   \{\mathbf{a},{\mathbf{b}},\mathbf{c}\}=\{\mathbf{c},{\mathbf{b}},\mathbf{a}\}.
\end{equation}
\item For arbitrary
$\mathbf{x},\mathbf{y},\mathbf{a},\mathbf{b}\in\mathbb{C}^4$,  the
following identity  holds
\begin{equation}\label{dbracket}
 [D(\mathbf{x},\mathbf{y}),D(\mathbf{a},\mathbf{b})]=
D(D(\mathbf{x},\mathbf{y})\mathbf{a},\mathbf{b})
-D(\mathbf{a},D(\mathbf{y},\mathbf{x})\mathbf{b}).
\end{equation}
 \end{enumerate}

Properties of the previous proposition are the defining properties
for the Jordan triple products associated with a homogeneous spaces,
see \cite{L77} and \cite{F04}. If the Euclidean inner product of
$\mathbb{C}^4$ is used  in the definition (\ref{tripleproddef}),
this triple product is the triple product of the bounded symmetric
domain of type IV in Cartan's classification, called the spin
factor. A similar triple product was obtained \cite{F04} for the
ball of relativistically admissible velocities under the action of
the conformal group. As we will see later, the geometric tri-product
(\ref{tripleproddef}) is useful in defining the action od the
Lorentz group on $\mathbb{C}^4$.

   The space $\mathbb{C}^4$
with a form (\ref{Minkbilin}) for the given metric tensor and a
geometric tri-product  (\ref{tripleproddef}) will be  denoted by
$\mathcal{S}^4$. As we have seen, the space $\mathcal{S}^4$ can be
used to represent the space-time coordinates and the relativistic
momentum variables. The form (\ref{Minkbilin}) on it defines both
the interval and the symplectic form and the tri-product may be used
to define the action of the Lorentz group. Thus, we propose to call
$\mathcal{S}^4$  the \textit{complex relativistic phase space}.

\section{The quasi-orthogonal group $QO ({\bf {\cal S}}^4)$ and its Lie algebra $qo ({\bf {\cal S}}^4)$}

As in  \cite{Hehl} we define:

 \noindent \textbf{Definition 7.1} An
invertible linear map $T\in GL({\bf {\cal S}}^4 )$ which preserves
the scalar product (\ref{Minkbilin}) will be called a
\textit{quasi-orthogonal} map. The group of all quasi-orthogonal
maps on ${\bf {\cal S}}^4 $ will be called the
\textit{quasi-orthogonal group} and denoted by $QO({\bf {\cal
S}}^4)$. The Lie algebra of $QO({\bf {\cal S}}^4)$ will be denoted
by $qo({\bf {\cal S}}^4).$

The  Lorentz group, preserving the intervals, also preserves the
scalar product (\ref{Minkbilin}). Thus, $QO({\bf {\cal S}}^4)$ can
be identified with the Lorentz group and its Lie algebra $qo({\bf
{\cal S}}^4)$ with the Lorentz algebra.

 From the definition of $QO({\bf {\cal S}}^4)$
we have
\begin{equation}\label{Dinvdef}
QO({\bf {\cal S}}^4)=\{g\in GL({\bf {\cal
S}}^4):<g\mathbf{a}|g\mathbf{b}>=<\mathbf{a}|\mathbf{b}>,\;\mathbf{a},\mathbf{b}\in
{\bf {\cal S}}^4\}.
\end{equation}
If $g(t)$ is a smooth curve in $QO({\bf {\cal S}}^4),$ with
$g(0)=I,$ the identity map on ${\bf {\cal S}}^4$, then $X:=g'(0)\in
qo({\bf {\cal S}}^4).$  Since $g(t)\in QO ({\bf {\cal S}}^4),$ from
(\ref{Dinvdef}) we have
\[<g(t)\mathbf{a}|g(t)\mathbf{b}>=<\mathbf{a}|\mathbf{b}>.\]
Differentiating this by $t$ and substituting $t=0$, we obtain
\begin{equation}\label{antisym}<X\mathbf{a}|\mathbf{b}>+<\mathbf{a}|X\mathbf{b}>=0.\end{equation}
In  basis $\{\mathbf{u}_\mu\}$ this equation takes the form
\[ (\eta _{\gamma\beta}{X _\alpha}^\gamma  +\eta
_{\gamma\alpha} {X _\beta}^\gamma)a^\alpha b^\beta=0,
\]
where the operator $X$ is represented by the mixed tensor ${X
_\alpha}^\gamma$. This mean that the matrix $X_{\alpha\beta} =\eta
_{\gamma\beta}{X _\alpha}^\gamma$ is antisymmetric. The space of
such antisymmetric two-tensors is a 6-dimensional complex space that
is denoted in the literature as $\mathcal{M}^6,$ see \cite{Hehl}.

 Using the triple product (\ref{tripleproddef}) on ${\bf {\cal S}}^4$ we see that the
linear operator
$D_{\alpha\beta}=\hat{D}(\mathbf{u}_\alpha,{\mathbf{u}}_\beta)$,
defined by (\ref{Dhatdef}) act on the basis elements as
\begin{equation}\label{Dalphabeta}
   D_{\alpha\beta}\mathbf{u}_\gamma=\hat{D}(\mathbf{u}_\alpha,{\mathbf{u}}_\beta)
\mathbf{u}_\gamma=-\eta_{\gamma\alpha}\mathbf{u}_\beta+\eta_{\beta\gamma}
\mathbf{u}_\alpha =-D_{\beta\alpha}\mathbf{u}_\gamma.
\end{equation}
 Thus $D_{\alpha\beta}$ are elements and span the Lie algebra
$qo({\bf {\cal S}}^4)$. Note that for $\alpha \neq\beta$,
$D_{\alpha\beta}={D}(\mathbf{u}_\alpha,{\mathbf{u}}_\beta)$.  We can
express this Lie algebra as
\begin{equation}\label{dinvspin}
qo({\bf {\cal S}}^4)=\{ x^{\alpha\beta}D_{\alpha\beta} :\;\:
x^{\alpha\beta} \in C,\;x^{\beta\alpha}=-x^{\alpha\beta} \}.
\end{equation}

The Lie bracket of the basis elements of this Lie algebra can be
calculated by use of (\ref{dbracket}) and (\ref{Dalphabeta}) as
\begin{equation}\label{brackeDab}
   [D_{\mu\nu},D_{\alpha\beta}]=D(D(\mathbf{u}_\mu,{\mathbf{u}}_\nu)\mathbf{u}_\alpha,{\mathbf{u}}_\beta)
 -D(\mathbf{u}_\alpha,D({\mathbf{u}}_\nu,\mathbf{u}_\mu){\mathbf{u}}_\beta)
\end{equation}
\[= \eta _{\nu \alpha} D_{\mu\beta}-
  \eta _{\mu\alpha} D_{\nu\beta} +\eta _{\nu \beta}D_{\alpha\mu}
- \eta _{\mu\beta}D_{\alpha\nu },\] which is similar to the
commutation relations (\ref{Lortzcommut}) of the angular momentum in
the Lorentz group.

\section{Spin 1 representation of the Poincar\'{e} algebra on ${\bf {\cal
S}}^4$}

To obtain a representation of the Poincar\'{e} algebra on ${\bf
{\cal S}}^4$ we will use the complex extension of the algebra
$\mathcal{L}$ defined by (\ref{LieAlgGradetReal}). The subspace
$\mathcal{L}_1$ becomes the relativistic phase space ${\bf {\cal
S}}^4$ and $\mathcal{L}_0=qo({\bf {\cal S}}^4)$. The symplectic
structure on $\mathcal{L}_1={\bf {\cal S}}^4$ defines a nontrivial
bracket which results in a new subspace $\mathcal{L}_2$ of grade 2
consisting of constants. The brackets of elements of $\mathcal{L}_2$
with any element of $\mathcal{L}$ are trivial. Thus, we define a
graded complex Lie algebra
\begin{equation}\label{Lieofspin}
   \mathcal{L}(\mathcal{S}^4) =\mathcal{L}_0\oplus\mathcal{L}_1\oplus\mathcal{L}_2,
\end{equation}
with $\mathcal{L}_1=\mathcal{S}^4,$ $\mathcal{L}_0=qo({\bf {\cal
S}}^4),$  defined by(\ref{dinvspin}) and $\mathcal{L}_2=\mathbb{C}$.

 The brackets on $\mathcal{L}_0$ are the
usual operator brackets. We define the bracket of any element of
$\mathcal{L}(\mathcal{S}^4)$ with any element of $\mathcal{L}_2$ to
be trivial. The bracket of two elements of $\mathcal{L}_1,$ defined
by (\ref{liebracL1}), is constant and can be considered as an
element of $\mathcal{L}_2$. Finally, the bracket  the basis elements
of $\mathcal{L}_0$ and $\mathcal{L}_1$, by use of
(\ref{Dalphabeta}), is defined as
\[[D_{\alpha\beta},\mathbf{u}_\mu]=\frac{1}{2}(\{\mathbf{u}_\alpha,{\mathbf{u}}_\beta,
   \mathbf{u}_\mu \}-\{\mathbf{u}_\mu,{\mathbf{u}}_\alpha,
   \mathbf{u}_ \beta\})\]
   \begin{equation}\label{bracketDandc}
=\frac{1}{2}(D_{\alpha\beta}\mathbf{u}_\mu
   -D_{\beta\alpha}\mathbf{u}_\mu)=D_{\alpha\beta}\mathbf{u}_\mu.
\end{equation}
Similarly,
 \[  [\mathbf{u}_\mu , D_{\alpha\beta}]=D_{\beta\alpha}\mathbf{u}_\mu=
  -D_{\alpha\beta}\mathbf{u}_\mu.\]

We define a representation of the Poincar\'{e} algebra in
$\mathcal{L}(\mathcal{S}^4)$ by
\begin{equation}\label{repsspin1}
 \pi(P_\alpha)=\mathbf{u}_\alpha ,\;\;\; \pi(M_{\alpha\beta})=
   D_{\alpha\beta}.
\end{equation}
From (\ref{liebracL1}) it follows that (\ref{comuttrans}) hold. By
use of (\ref{Dalphabeta}) we get
\begin{equation}\label{DMink}
   [\pi(M_{\alpha\beta}), \pi(P_\mu)]=
 D_{\alpha\beta}\mathbf{u}_\mu= \eta _{\mu\beta}\mathbf{u}_\alpha-\eta _{\mu\alpha}\mathbf{u}_\beta
    =\eta _{\mu\beta}\pi(P_\alpha)-\eta _{\mu\alpha}\pi(P_\beta)
\end{equation}
and (\ref{comutangtrans}) holds. By use of (\ref{brackeDab})
  we get
\[ [ \pi(M_{\mu\nu}),\pi(M_{\alpha\beta})]=
[D_{\mu\nu},D_{\alpha\beta}]\]
\[=\eta _{\mu\beta}\pi(M_{\nu\alpha})+\eta _{\nu\alpha}
 \pi(M_{\mu\beta})-\eta _{\mu\alpha}\pi(M_{\nu\beta})
 -\eta _{\nu\beta}\pi(M_{\mu\alpha}).\]
which coincides with (\ref{Lortzcommut}). Thus, $\pi$ defined by
(\ref{repsspin1}) defines a representation of the Poincar\'{e}
algebra into $\mathcal{L}(\mathcal{S}^4)$.

 Under this representation, the generators of the
boosts are represented as $
  \pi(M_{0 \beta})=
  D_{0\beta}$ for  $\beta \in \{1,2,3\}$.
The representation of the boosts is given by  the exponent of
$D_{0\beta}$ on $M$ and $iM.$  Since $D_{0\beta}^3=D_{0\beta}$, the
matrix of the boost in $x$ direction ($\beta =1$) is
\begin{equation}\label{bostreprpis4}
 \exp (\varphi\pi(M_{0 1}))=\left(\begin{array}{cccc}
\cosh \varphi&\sinh \varphi&0&0\\
\sinh \varphi&\cosh \varphi&0&0\\
0&0&1&0\\
0&0&0&1
\end{array}\right),
\end{equation}
which is the usual \textit{Lorentz  four-momentum and space-time
transformation} for the boost in the $x$-direction, where
$\tanh\varphi=v/c,$ and $\mathbf{v}=(v,0,0)$ is the relative
velocity between the systems.

Similarly, the generators of the rotation are represented as
$\pi(M_{\alpha\beta})=D_{\alpha\beta}$ for $ \alpha,\beta \in
\{1,2,3\}$. Since in this case $D_{\alpha\beta}^3=-D_{\alpha\beta},$
their exponent defines the regular rotations on the subspaces $M$
and $iM$. Thus, the representation $\pi $ is a spin 1
representation. Note that both subspaces $M$ and $iM$ are invariant
under this representation.

It is known that an electric field act as a generator of a boost and
a magnetic field as a generator of rotations on the four-momentum of
a charged particle. The four momentum is represented by the real
part of the space ${\bf {\cal S}}^4.$ So, any electromagnetic field
strength tensor could be identified with an element of the real part
of the space $qo({\bf {\cal S}}^4)$  as $
\mathfrak{F}=F^{\alpha\beta} D_{\alpha\beta}.$  Under this
representation, the electric field is presented by
$\mathbf{E}=F^{0j} D_{0j}$ with $j=1,2,3$  and the magnetic field is
represented by $\mathbf{B}=B_j
\frac{1}{2}{\epsilon_{0j}}^{\alpha\beta}D_{\alpha\beta}=B^1
D_{23}+B^2 D_{31}+B^3 D_{12},$ with ${\varepsilon_{0j}}^{\mu\nu}$
being the Levi-Civita symbol. The Lorentz group will act properly on
 the electric and magnetic components of the field. The Lorentz
 force on a particle with four-momentum $\mathbf{p}$ is given by
  $[\mathfrak{F},\mathbf{p}]=\mathfrak{F}(\mathbf{p})$.
According \cite{Hehl} the imaginary part of $\mathcal{M}^6$ can be
identified with the excitation of the field.

Sometimes it is useful to represent the electromagnetic field not as
the  electromagnetic tensor $ \mathbf{F}$ in which the electric and
magnetic components are linearly independent, but as a complex
Faraday vector $\mathbf{F}_c=\mathbf{E}+i\mathbf{B}$ in which both
components are complex dependent vectors. As we will show is the
next section, this can be obtained by by use of a spin 1/2
representation the Poincar\'{e} algebra into
$\mathcal{L}(\mathcal{S}^4)$.

\section{Spin 1/2 representation of the Poincar\'{e} algebra on ${\bf {\cal
S}}^4$}

 The complex linear space $qo({\bf {\cal S}}^4)$ is a subspace of the
space of operators on ${\bf {\cal S}}^4.$ From (\ref{Dalphabeta}) it
follows that $(D_{\alpha\beta})^3=-\eta _{\alpha\alpha}\eta
_{\beta\beta}D_{\alpha\beta}\in qo({\bf {\cal S}}^4)$ for any
distinct $\alpha,\beta$. Thus, $D_{0\alpha}^3=D_{0\alpha}$. Such
operators are called a tripotents. If $j,k\in\{1,2,3\}$, then then
$D_{j,k}^3=-D_{j,k}$ and $iD_{j,k}$ is a tripotent in this case.

Define now
$D_{0j}^\perp=\frac{1}{2}{\varepsilon_{0j}}^{\mu\nu}D_{\mu\nu},$
were ${\varepsilon_{0j}}^{\mu\nu}$ denotes the Levi-Civita symbol
with $\varepsilon ^{0123}=1$. For example $D_{01}^\perp
=D_{23},\;D_{23}^\perp =-D_{01}.$ With this definition
$iD_{0j}^\perp$ is a tripotent and $iD_{0j}^\perp
D_{0j}=D_{0j}iD_{0j}^\perp =0$.

  From this one gets that
\begin{equation}\label{ualphabeta}
 D^\pm_{0j}=D_{0j}\pm iD_{0j}^\perp
=D_{0j}\pm i\frac{1}{2}{\varepsilon_{0j}}^{\mu\nu}D_{\mu\nu}
\end{equation}
are  tripotents for any $j\in\{1,2,3\}$. By direct verification you
get
\[(D^\pm_{0j})^2=I, \;\;D^\pm_{0j}D^\pm_{0k}+
D^\pm_{0k}D^\pm_{0j}=0,\] which can be rewritten as the
\textit{canonical anticommutation relations}
\begin{equation}\label{CAR}
    \frac{1}{2}(D^\pm_{0j}D^\pm_{0k}+
D^\pm_{0k}D^\pm_{0j})=\delta_{jk} I.
\end{equation}

We introduce now another representation, which will be denoted by
$\pi^+$, of the Poincar\'{e} algebra in the graded Lie algebra
$\mathcal{L}(\mathcal{S}^4)$, defined by (\ref{Lieofspin}). We
represent first the the relativistic angular momentum by elements of
$\mathcal{L}_0$ as
     \begin{equation}\label{piplusangmoment}
\pi^+(M_{0j})=\frac{1}{2}D^+_{0j},\;\;\pi^+(M_{kl})=
\frac{i}{2}{\varepsilon_{kl}}^{0j}D^+_{0j}\end{equation} where
$D^+_{0j}$ is defined by (\ref{ualphabeta}) and $j,k\in\{1,2,3\}$.
For example, $\pi^+(M_{01})=\frac{1}{2}(D_{01}+ iD_{23})$ and
$\pi^+(M_{23})=\frac{1}{2}(D_{23}- iD_{01})=-i\pi^+(M_{01}).$ Also
for this representation the bracket on $\mathcal{L}_0$ is the usual
operator bracket.

To verify that this is a representation of the Lorentz algebra we
have to check that (\ref{Lortzcommut}) is satisfied. Because of the
symmetry in (\ref{Lortzcommut}), it is enough to check 4 cases of
these relations:
\[ [\pi^+(M_{23}),\pi^+(M_{12})]=\frac{1}{4}[(-iD_{01}+D_{23}),(-iD_{03}+D_{12})]\]
\[=\frac{1}{4}([iD_{01},iD_{03}]-[D_{23},iD_{03}]-[iD_{01},D_{12}]+[D_{23},D_{12}])\]
\[=\frac{1}{4}(-D_{31}+iD_{02}+iD_{02}-D_{31})=\frac{1}{2}(iD_{02}-D_{31})=-\pi^+(M_{31}),\]
\[[\pi^+(M_{01}),\pi^+(M_{31})]=[i\pi^+(M_{23}),\pi^+(M_{31})]
=i\pi^+(M_{21})=-\pi^+(M_{03})],\]
\[[\pi^+(M_{01}),\pi^+(M_{03})]=[i\pi^+(M_{23}),i\pi^+(M_{12})]\]
\[=-[\pi^+(M_{23}),\pi^+(M_{12})]=-\pi^+(M_{31}),\]
and
\[[\pi^+(M_{01}),\pi^+(M_{23})]=[\pi^+(M_{01}),-i\pi^+(M_{01})]=0.\]
 Thus, (\ref{Lortzcommut}) is
satisfied.

We represent the generators of translation, as for the
representation $\pi$, by $\pi^+(P_\mu)=\mathbf{u}_\mu \in
\mathcal{L}_1$, but we need to modify the bracket between
$\mathcal{L}_0$ and $\mathcal{L}_1$ (instead of (\ref{bracketDandc})
) to be
\[[D_{\alpha\beta}^+,\mathbf{u}_\mu]=(D_{\alpha\beta}^+ +\overline{D_{\alpha\beta}^+})\mathbf{u}_\mu
=2D_{\alpha\beta}\mathbf{u}_\mu .\] This imply that
$[\pi^+(M_{\alpha\beta}),\mathbf{u}_\mu]=D_{\alpha\beta}\mathbf{u}_\mu$
and from (\ref{DMink}) follow that commutation relations
(\ref{comutangtrans}) are satisfied for representation $\pi^+$.
 This finishes the proof that $\pi^+$  is a representation of the Poincar\'{e} algebra.

From (\ref{CAR}) follows that the operator $\pi^+(M_{jk})$, for
$j,k\in \{1,2,3\},$ representing angular momentum satisfies
$\pi^+(M_{jk})^2=-\frac{1}{4}I$. Thus, the flow generated by them is
\[\exp(\varphi\pi^+(M_{jk}))=\cos(\frac{1}{2}\varphi)I+\sin(\frac{1}{2}\varphi)\pi^+(M_{jk}),\]
showing that the representation  $\pi^+$ is a spin 1/2
representation. The operator $\pi^+(M_{0k})$ representing generators
of boosts satisfies $\pi^+(M_{jk})^2=\frac{1}{4}I$ and thus its flow
is given by
\[\exp(\varphi\pi^+(M_{0k}))=\cosh(\frac{1}{2}\varphi)I+\sinh(\frac{1}{2}\varphi)\pi^+(M_{0k}).\]

In addition to the representation $\pi^+$ we can define also a
representation $\pi^-$ by
\[\pi^-(M_{0j})=\frac{1}{2}D^-_{0j},\;\pi^-(M_{jk})=
\frac{i}{2}{\varepsilon_{jk}}^{\mu\nu}D^-_{\mu\nu}.\] This
representation is also a spin 1/2 representation of the Poincar\'{e}
algebra.

 Let us choose the  Newman-Penrose basis
$(\mathbf{l},\mathbf{m},\mathbf{n},\overline{\mathbf{m}})$, defined
in ${\bf {\cal S}}^4$ by ( see \cite{PenroseRindler} )
\[\mathbf{l}=\frac{1}{\sqrt{2}}(\mathbf{u}_0+\mathbf{u}_3),\;\;\mathbf{m}=\frac{1}{\sqrt{2}}(\mathbf{u}_1+i\mathbf{u}_2)\]
\begin{equation}\label{neumanPenroseBasis}
  \overline{\mathbf{m}}=\frac{1}{\sqrt{2}}(\mathbf{u}_1-i\mathbf{u}_2),\;\;\mathbf{n}=\frac{1}{\sqrt{2}}(\mathbf{u}_0-\mathbf{u}_3).
\end{equation}
 Direct calculation show that in this basis
the matrices of the generators of boosts $M_{0j}$ and the generators
of rotations $J_j=\frac{1}{2}{\varepsilon_j}^{kl}M_{kl}$ will have a
block-matrix form
\begin{equation}\label{spin1/2Pauli}
    \pi^+(M_{0j})=-\frac{1}{2}\left(
                               \begin{array}{cc}
                                 \bar{\sigma}_j & 0 \\
                                 0 & \sigma_j \\
                               \end{array}
                             \right),\;\;
                             \pi^+(J_{j})=\frac{i}{2}\left(
                               \begin{array}{cc}
                                 \bar{\sigma}_j & 0 \\
                                 0 & {\sigma_j }\\
                               \end{array}
                             \right),
\end{equation}
in which $\sigma_j$ are the Pauli matrices. Thus, under this
representation, the relativistic phase space ${\bf {\cal S}}^4$ with
the NP basis on it transforms as the Dirac bi-spinors.

As in the previous section, we use the connection of the
electro-magnetic field with relativistic angular momentum to
represent this field tensor by an elements of $\mathcal{L}_0=qo({\bf
{\cal S}}^4)$. We use now the complex Faraday vector
$\mathbf{F}_c=\mathbf{E}+i\mathbf{B}$ to represent the field and
represent it by use of the representation $\pi ^+$ as a
\textit{complex Faraday  tensor} defined as
\begin{equation}\label{complex electromagnetic tensor}
  \mathfrak{F}_c=F_c^j \pi^+(M_{0j}).
\end{equation}
A similar complex Faraday  tensor was introduced along ago by L.
Silberstein \cite{Silberstein} and was used later in \cite{Weiess}.
Note that the electromagnetic and the complex Faraday  tensor are
related as $\mathfrak{F}=\mathfrak{F}_c+\overline{\mathfrak{F}}_c.$

This representation could be useful for example in calculating the
evolution of momentum $\mathbf{p}(\tau)$ a charged particle in a
uniform electro-magnetic field. The evolution equation is
$\frac{d\mathbf{p}(\tau)}{d\tau}=\mathfrak{F}\mathbf{p}(\tau),$
which can be rewritten as $\frac{d\mathbf{p}(\tau)}{d\tau}=
(\mathfrak{F}_c+\overline{\mathfrak{F}}_c)\mathbf{p}(\tau).$ Using
the fact that the operators $\mathfrak{F}_c$ and
$\overline{\mathfrak{F}}_c$ commute, the solution of the evolution
equation is
\[\mathbf{p}(\tau)=e^{\mathfrak{F}\tau}\mathbf{p}_0=
e^{\overline{\mathfrak{F}}_c\tau}e^{\mathfrak{F}_c\tau}\mathbf{p}_0,\]
where $\mathbf{p}_0=\mathbf{p}(0).$

We denote the complex Lorentz invariant associated with the
electro-magnetic field by $z(F)=\mathbf{F}_c^2.$ From (\ref{CAR})
follows that $\mathfrak{F}_c^2=(\frac{\sum
(F_c^j)^2}{4})I=\frac{z(F)}{4}I.$ Define now a complex number
$w^2=\frac{z(F)}{4}$. With this definition $\mathfrak{F}_c^2=w^2I$
and
\[e^{\mathfrak{F}_c\tau}=\sum\mathfrak{F}_c^n\frac{\tau^n}{n!}=
\cosh (w\tau) I +\frac{\sinh (w\tau)}{w}\mathfrak{F}_c.
\]
This give an explicit solution of the evolution equation.

\section{Conclusions and Discussions}

In this paper we introduced  a complex relativistic phase space
$\mathcal{S}^4$ as the space $\mathbb{C}^4$ with a scalar product
(\ref{Minkbilin}), based on the relativistic metric tensor, and with
a geometric tri-product (\ref{tripleproddef}) on it. We have seen
that the space $\mathcal{S}^4$ can be used to represent the
space-time coordinates and the relativistic momentum variables. The
scalar product  defines both the Lorentz scalar product and a
relativistic extension of the symplectic form. We have shown that
the Lorentz algebra is represented by natural operators of the
tri-product.

We constructed both spin 1 (in chapter 5) and spin 1/2 (in chapter
6) representations of the Poincar\'{e} group by natural operators of
the tri-product on the phase space. For the spin 1 representation,
the generators of boosts are presented by operators with the meaning
of electric field strength tensors while the generators of rotations
are presented by operators with the meaning of magnetic field
strength tensors. For the spin 1/2 representation, the generators of
boosts and  of rotations are presented by the complex Faraday
electromagnetic tensor. We have shown that if we use the
Newman-Penrose basis on the phase space, we obtain the Dirac
bi-spinors under the spin 1/2 representation.

We want to propose an explanation of the fact that use of the real
electromagnetic tensor for representation of the relativistic
angular momentum led to a spin 1 representation, while the use of
the complex Faraday tensor led to a spin 1/2 representation and the
Dirac bi-spinors.

The last representation is somehow related with an action of an
electro-magnetic field on an electron. An electron, in addition of
being a charged particle, has a magnetic momentum, called the spin.
A real electromagnetic tensor account only on action of a field on
the charged particle, but do not take in account the spin
precession. We conject that the complex Faraday tensor describe the
full action of the electromagnetic field on the electron. This may
explain why the solution of the evolution equation of a charge in a
constant uniform field is significantly simpler if we use the
complex Faraday tensor instead of the electromagnetic tensor.

\medskip

I want to thank Gerald Kaiser and Yakov Itin for helpful remarks and
suggestions.

\end{document}